\documentclass[12pt,preprint]{aastex}

\slugcomment{Version 1.10}
\shorttitle{Imaging Atmospheric Cherenkov Telescope Arrays as Stellar Intensity Interferometers}
\shortauthors{Le~Bohec and Holder }
\begin{document}

\title{Optical Intensity Interferometry with Atmospheric Cherenkov Telescope Arrays}

\author{S. Le Bohec}
\affil{Department of Physics, University of Utah,
               Salt-Lake-City, UT, 84112-0830, USA}
\email{lebohec@physics.utah.edu}
\and       
\author{J. Holder}
\affil{School of Physics and Astronomy, University of Leeds, Woodhouse Lane,
  Leeds, LS2 9JT, UK}
\email{jh@ast.leeds.ac.uk}

\begin{abstract}

In the 1970s, the Narrabri intensity interferometer was used to measure 32
stellar diameters; some as small as 0.4 milli-arc-seconds (mas). The
interferometer consisted of a pair of 6.5\,m telescopes ($\rm 30\,m^2$) with
relatively crude optics, similar to those currently in use as Atmospheric
Cherenkov Telescopes (ACT). We explore the possibility of implementing a
modern intensity interferometer on an ACT array. Developments in fast digital
signal processing technology now make such a system relatively easy to 
implement, and
provide improved sensitivity. Allowing measurements at short wavelength ($\rm
<400\,nm$), with long baselines ($\rm > 100\,m$), which are still challenging for
Michelson interferometers, present ACT arrays could be used to probe angular
structures as small as $\rm \sim 0.2\,mas$, and smaller with large array
projects already being discussed. This would provide measurements of stellar
diameters, binary systems, circumstellar environments and, possibly, stellar
surface features. ACT arrays could be used as intensity interferometers
during bright moon periods, providing valuable scientific output for
little expense and no impact on the $\gamma$-ray observing schedule.

\end{abstract}


\section{Introduction}
The first successful measurement of the angular diameter of a star was
made using the Michelson optical interferometer on Mount Wilson, over a
baseline of 20 feet, \cite{michelson, michelsonpease}. Difficulties
encountered in the subsequent operation of a 50 foot baseline
interferometer brought the development of stellar interferometry to a
halt for thirty years. No further progress was made until the sixties,
with the construction and operation of the Narrabri intensity
interferometer.

Intensity interferometry was originally developed in the early fifties in the
field of radio astronomy by R.~Hanbury~Brown, who soon extended the idea to
the optical domain. After several laboratory tests and actual stellar
intensity interferometer prototypes, the technique culminated in the 60s with
the interferometer constructed under the direction of Hanbury Brown and Twiss
in Narrabri. The Narrabri intensity interferometer was successfully used to
study 32 stars \cite{brown1974}, all brighter than B=+2.5, providing angular 
diameter measurements as small as $0.41\pm0.03$
milli-arc-seconds (mas). Despite these successes, and plans for a larger
instrument, the technique was then abandoned to the benefit of Michelson
interferometry, which had become attractive again because of technological
developments in photo-detection and optics. Since then, to our knowledge,
intensity interferometry has not been used for astronomical
measurements.

In this paper, we revisit intensity interferometry and the scientific potential
it could offer if implemented on large Atmospheric Cherenkov Telescopes using
modern signal processing technology. In section 2, we briefly review the
principles of intensity interferometry. In section 3, we describe how an
intensity interferometer could be implemented with an ACT array and in the two
following sections, we comment on some specific points regarding the 
telescope's optics and the electronics. In the last section, we conclude and 
comment on the possible future developments and scientific potentials of 
intensity interferometry.

\section{Principles of intensity interferometry}
\subsection{ Intensity interferometry}

Ideas in this section are for the greatest part taken from the excellent
monograph by R. Hanbury Brown \cite{rhbbook} in which he describes the theory
of intensity interferometry and the technical aspects of the construction and
operation of the Narrabri interferometer. Intensity interferometry is rather
counter-intuitive when applied to the optical domain, and the idea was first
received with controversy until several experimental tests confirmed the
technique was well-founded. Today it is even applied to $\rm \pi_0$'s emerging
from nuclear collisions \cite{boal}, for synchrotron X-ray beam diagnostics
\cite{yang1994,tai2000} and in other fields. 

In a stellar intensity interferometer, the light from a star is received by
two separated photoelectric detectors through a  narrow optical bandwidth
filter. Intensity interferometry relies on the fact that the fluctuations in
the current output by the two detectors ($\rm \Delta i_1$ and $\rm \Delta
i_2$) are partially correlated.  The principal component of the fluctuation is
the classical shot noise which does not show any correlation between the two
detectors. In addition, there is a smaller component, the wave noise,
which can be seen as the beating between the different Fourier components of the
light reaching the detectors.  The wave noise shows correlation
between the two detectors provided there is some degree of coherence between
the light at the two detectors. The important point here is that the
correlation is a function of the difference in phase between the low frequency
beats at the two detectors. This correlation does not depend on the phase
difference of the light at the two detectors. The requirements on the
mechanics and optics of an intensity interferometer are therefore much less
stringent than in the case of a Michelson interferometer. Because of this
correlation, the time integrated product of the current fluctuations $\rm
\Delta i_1$ and $\rm \Delta i_2$ is positive and provides a measurement of the
square of the degree of coherence $\rm \gamma_d$ of the light at the two
detectors. 

\begin{figure}[function]
\epsscale{.40}
\plotone{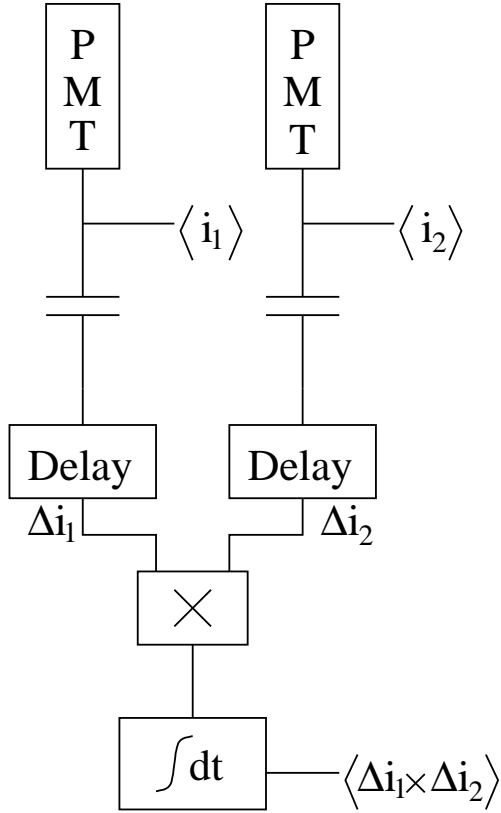}
\caption{In each receiver, a photo-multiplier tube is used to measure the star
  lightintensity. The DC components $\rm \langle i_1\rangle$ and $\rm \langle i_2\rangle$ of both signals are measured. The AC 
  components $\rm \Delta i_1(t)$ and $\rm \Delta i_2(t)$ must be placed in time before being multiplied together in the 
  correlator. This product is then integrated over time, providing a
  measurement of the degree of coherence at the two receivers.}
\label{function}
\end{figure}

\begin{figure}[ang_diam_veritas]
\epsscale{.80}
\plotone{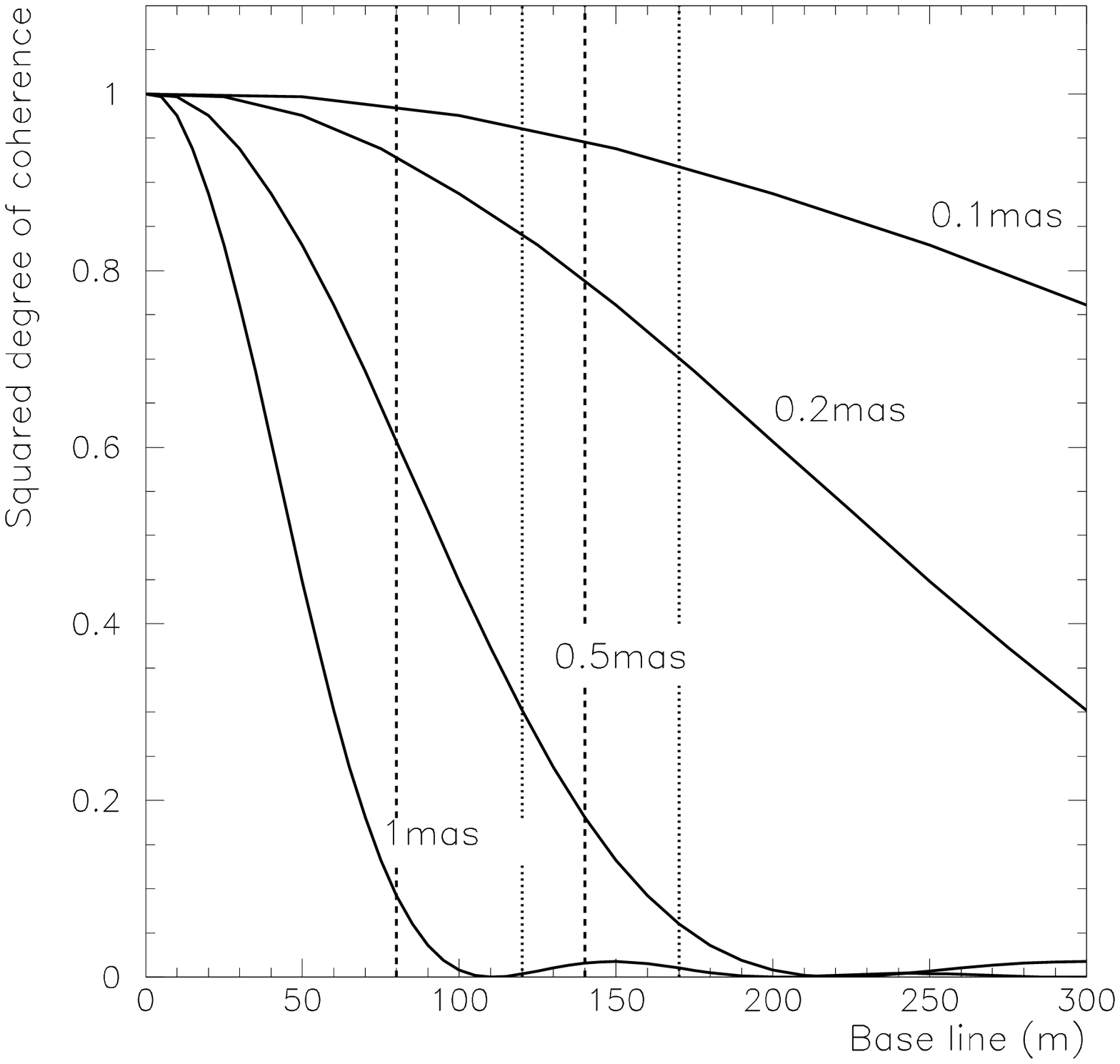}
\caption{
  The squared modulus of the degree of coherence of the light at two telescopes
  as a function of the baseline for different stellar diameters. The dashed 
  and dotted lines indicate the baselines at zenith in VERITAS-4 and HESS-4
  respectively.}
\label{ang_diam_veritas}
\end{figure}

Figure \ref{function} is a functional schematic of an intensity
interferometer. It can be shown that $\rm \langle \Delta i_1(t)\Delta
i_2(t)\rangle /(\langle i_1\rangle\langle i_2\rangle))=|\gamma_d|^2$.  The
degree of coherence is equivalent to the fringe visibility measured with a
Michelson interferometer. In the case of a star modeled as a uniform disk, it
depends upon the star's angular diameter $\rm \theta$, the wavelength, $\rm
\lambda$, and the distance between the two detectors, $\rm d$, reaching zero
when $\rm \rm d=1.22\lambda/\theta$. More generally, according to the van
Citter-Zernike theorem, $\rm \rm \gamma_d$, the complex degree of coherence,
is the normalized Fourier transform of the source intensity distribution
projected on a line parallel to the line joining the two detectors.  Figure
\ref{ang_diam_veritas} shows the squared degree of coherence as a function of
the base-line as it would be measured by an ideal intensity interferometer
observing uniform-disk shaped stars of different angular diameters. Baselines
available in existing ACT arrays would permit to probe the stellar diameter
range from $\rm \sim 0.1\,mas$ to $\rm \sim 1\,mas$.  A $\rm 1\,km$ baseline,
as might be offered by the next generation of ACT arrays, would make possible
the measurement of stellar diameters smaller than $\rm 50\,\mu as$.

\subsection{Noise considerations}
The shot noise from both channels is responsible for most of the multiplier
output fluctuations to which the wave noise correlation has to be compared
for the sensitivity of a specific experiment to be estimated. Hanbury
Brown showed that the signal to noise ratio can be expressed by 
equation \ref{signoise}.
\begin{equation}
\rm (S/N)_{RMS}=A\alpha n |\gamma_d|^2 (\Delta fT/2)^{1/2}
\label{signoise}
\end{equation}
where A is the collection area of each telescope, $\alpha$ the photo-detector 
quantum efficiency, $\rm \Delta f$ the bandwidth of the electronics including
the photo-detector, T the integration time and n the intensity of the source in
photons per unit optical bandwidth, per unit area and per unit time.  Using a 
pair of $\rm 100\,m^2$ telescopes with 30\% quantum efficiency photo-detectors, 
1\,GHz electronics during a full 5 hours night would permit to measure the 
diameter of stars with visual magnitude 6.7 with 5 standard deviations at 
$|\gamma_d|^2 =0.5$. This would result in a measurement of the diameter with
an accuracy of 14\%. The diameter of a magnitude 5 star could be measured with
an accuracy of 3\% in the same amount of time.

\subsection{ Comparison with Michelson Interferometry}

After decades of development, a number of world-class instruments for
long-baseline optical interferometry are currently operating or soon to come
online - for a review see \citep{monnier}. These are, without exception, based
upon the Michelson technique for interferometry. The advantage of this method
is clear. As it relies on the visibility of fringes produced by the amplitude
interferences formed by the light collected by two telescopes, it permits to
measure stars much dimmer than intensity interferometry with same size
telescopes. 

Michelson interferometry is also a very challenging 
technique as the relative length of the paths of light have to be controlled 
with an accuracy smaller than the wavelength of the light being measured. 
This requires high optical quality and high precision tracking. The situation 
is further complicated by the effects of atmospheric turbulence which have to 
be actively compensated for. These difficulties have constrained Michelson 
interferometry to smaller baselines (with the exception of CHARA \cite{chara1}
all 
interferometers have maximal  operational baselines of less than $\rm100\,m$) 
and longer wavelengths ( most interferometers work in the near-IR and mid-IR) 
while the angular resolution is proportional to the baseline and to the inverse of the wavelength. 
 
Intensity interferometry, however, only requires control of the path of light
to an accuracy fixed by the bandwidth of the electronics. For a system with
$\rm 1\,GHz$ bandwidth, differences in the light paths of a few centimeters
would affect the correlation level by less than 10\%. 
It is also therefore essentially insensitive to atmospheric fluctuations and
the use of shorter wavelengths does not result in any extra
difficulty. Intensity interferometry also permits to simultaneously measure
the degree of coherence between any two telescope pair of an array, while with
Michelson interferometry this is impossible without a loss in sensitivity.  A
two-telescope based intensity interferometer does not provide the phase of the
complex degree of coherence and an image can only be reconstructed up to a
central symmetry. With a system consisting of three or more telescopes,
however, the phase can be reconstructed \cite{aviv2,gamo}. Intensity
interferometry, as with Michelson interferometry, allows
model-independent reconstruction of the object image. The major drawback
associated with intensity interferometry is the demand for very large
quantities of light. The necessary large light collectors do not, however, need
to offer optical astronomy quality as they are only required to
concentrate the light on a photo-detector. Furthermore, such light collectors
are readily available in ACT arrays.

\section{Possible Implementations}

\subsection{The Narrabri interferometer}
We begin this section with a brief description of the Narrabri intensity
interferometer, in order to have an experiment of measured sensitivity with
which to compare.

The Narrabri interferometer consisted of two telescopes, each 6.5\,m in
diameter, with an 11\,m focal length. The telescopes were carried on trucks
running on a circular railway track 188\,m in diameter. This allowed the
interferometer to operate with a baseline of 10\,m to 188\,m, and to track any
star while keeping the line joining the two telescopes perpendicular to the
direction of the star. Having movable telescopes removes the need for
additional delay lines to bring the signals in time. At the focus of each
telescope, the converging light was collimated and passed through an
interference filter centered on 443\,nm with a passing band width of 10\,nm. The
light was then focused onto a photomultiplier tube which converted the light
intensity to a current with  25\% quantum efficiency at 440\,nm and a 60\,MHz
effective bandwidth. The signals were sent to the control building where the
correlator was located.

The DC component of the signals was measured and recorded while the AC
components were sent to the input of the correlator, a 4 transistor based
linear multiplier. The DC component to be measured on the output of the
multiplier is very small in comparison to the random fluctuations ($\rm \sim
100\,dB$). This difficulty was resolved by integrating the output of the
multiplier over $\rm 100\,s$ time intervals. The drifting effects of small
offsets that would have drowned the signal also had to be eliminated. This was
achieved by applying the technique of phase-switching, which consisted of
inverting the phase of the signals on the input of the multiplier at rates of
10\,kHz and 10\,Hz respectively. On the output of the multiplier, these
switching frequencies were amplified and demodulated before the signal was
integrated, thus removing any significant contamination due to offsets.  This clearly was the
critical part of the experiment and great efforts had to be made in order to
reduce spurious correlations to acceptable levels. For an unresolved
($\gamma_d=0$) 0 visual magnitude star observed during one hour, the predicted
signal to noise ratio amounts to $\rm \sim 130$. The actual signal to noise
ratio was degraded by a factor 5 because of optical losses and excess noise of
various origins. This limited observations to stars brighter than $\rm
B=+2.5$.

Hanbury Brown and Twiss proposed several possible improvements to attain
higher sensitivities. Aside
from increasing the telescope size and improving the electronics bandwidth,
they proposed to multiply the number of independent optical channels used in
the measurements. The sensitivity does not depend on the optical band-width,
and so this can be easily achieved by simultaneously using several narrow
optical bands. Sensitivity improvements can also result from observing the
same star with an array of telescopes providing measurements over several
baselines simultaneously \cite{herrero}. In a N telescope array, the number 
of baselines is ${N(N-1)} \over {2}$. These
ideas could be combined with the technological developments during the last 30
years which have provided higher bandwidth photo-detectors and electronics and
the possibility of processing digitized signals at high speed.

\subsection{Ground-based $\gamma$-ray telescopes and intensity interferometry}
The two telescopes of the Narrabri interferometer were also used in a search
for astronomical sources of very high energy ($\rm E>300GeV$)
$\gamma$-rays\cite{grindlay75, grindlay75cena}. Cosmic rays and $\gamma$ rays
reaching the Earth's atmosphere trigger extensive particle showers which
produce Cherenkov light. This light can be detected with large telescopes
equipped with fast photo-detectors and electronics. Since these early
measurements, the field of ground based $\gamma$-ray astronomy has matured and
more than 30 very high energy $\gamma$ ray emitters are now known\cite{aharonian2005}. This
success can be entirely attributed to the sensitivity achieved with imaging
atmospheric Cherenkov telescopes \cite{weekes} used in stereoscopic
arrays. Existing major ground based $\gamma$-ray observatories
like CANGAROO\cite{kubo}, H.E.S.S.\cite{bernlhor, cornils}, MAGIC\cite{lorenz} and
VERITAS\cite{krennrich} consist of arrays of ACTs which satisfy many of the
specifications for a productive intensity interferometer.  We summarize the 
main characteristics of existing and projected ACT arrays and their
corresponding capabilities as intensity interferometers in Table
\ref{chertel}. Existing ACT arrays typically extend over 200\,m, making them
comparable to the Narrabri interferometer from the point of view of the
angular resolution they could achieve. Telescopes have diameters ranging from 
10\,m to 17\,m providing a gain of 2.8 to 8.0 in sensitivity (0.94 to 2.1 
magnitude) when compared to the Narrabri interferometer.  Furthermore, these 
arrays (except for MAGIC) consist of 4 telescopes, and could permit
measurements along up to six baselines simultaneously. This corresponds to a 
sensitivity gain of $\sim2.5$, or one 
magnitude. Table \ref{chertel} also provide estimates of the sensitivity that
could be achieved with these arrays. Sensitivity estimates resulting from 
equation \ref{signoise} are probably optimistic as some loss is unavoidable. 
On the other hand, the simple scaling from the Narrabri performances does not 
account for the further sensitivity improvement that must result from the 
tremendous progress in signal processing electronics since the
60s. Furthermore, Ofir and  Ribak \cite{aviv1} have recently shown that with 
a many telescope array, by using higher order correlation, it might be
possible to more than double the signal to noise ratio and therefore improve 
sensitivity by almost one magnitude. Even without considering this, the table 
clearly indicates the high potential 
offered by ACT arrays in the domain of stellar interferometry: it shows the 
possibility of baselines to wavelength ratios never attained before with 
sensitivity comparable to modern Michelson interferometers.

\begin{table}
\begin{center}
\begin{tabular}{cccccccc}
\tableline\tableline
Array          & N & A     & n    & $\rm d_{min} - d_{max}$
& $\rm \theta_{min} - \theta_{max} $ & $V_{Max}$ & $T_{V=5}$\\
\tableline
MAGIC-II          & 2 & 227   & 1    & $\rm  85        $   &  1.2         & 4.7 (6.0) & 0.16\\
CANGAROO       & 4 &  57   & 6    & $\rm 100~to~184 $   &  0.5 to 1.0  & 4.2 (5.5) & 2.61\\
HESS-4         & 4 & 108   & 6    & $\rm 120~to~170 $   &  0.6 to 0.8  & 4.9 (6.1) & 0.73\\
VERITAS-4      & 4 & 113   & 6    & $\rm 80~to~140  $   &  0.7 to 1.2  & 4.9 (6.2) & 0.66\\
VERITAS-7(*)   & 7 & 113   & 21   & $\rm 80~to~160  $   &  0.6 to 1.2  & 5.6 (6.9) & 0.66\\
HESS-16(*)     &16 & 108   & 120  & $\rm 120~to~510  $  &  0.2 to 0.8  & 6.5 (7.8) & 0.73\\
Next-Generation (**)&50 & 100   & 1225 & $\rm 80~to~1000  $  &  0.1 to 1.2  & 7.7 (9.0) & 0.85\\
\tableline
Narrabri       & 2 & 30    & 1    & $\rm 10m \to 188m $ &  0.5 to 10   & 2.5 (3.8) & 9.4\\
\tableline
\end{tabular}
\end{center}
\caption{$\rm N$
represents the number of telescopes, $\rm A$ is the light collection area of
each telescope in $\rm m^2$ (all of these arrays consist of identical
telescopes), $\rm n$ is the number of baselines simultaneously available, $\rm
d_{min} - d_{max}$ indicates the range of baselines in meters for observation
at zenith. The corresponding range of angular diameters in mas ($\rm 1.22
\lambda / d $) for observations at $\rm 400\,nm$ is indicated by $\rm
\theta_{min} - \theta_{max} $. $V_{Max}$, is the largest stellar
magnitude that could be attained with such an array. The first number was
obtained simply by scaling the largest magnitude measured with the Narrabri
interferometer by the increase in light collection area. The number between
brackets was obtained by applying equation \ref{signoise} with a quantum
efficiency of $25\%$, a bandwidth of 100\,MHz, and an observing time of 1 hour
to detect the non-resolved correlation ($ |\gamma_d|^2=1 $) with 5 standard
deviations. Both numbers are scaled to account for the number of simultaneously
available baselines under the assumption that the star is not resolved even with
the largest baseline. $T_{V=5}$ (in hours) is the time required to reach a
 five standard deviation detection of the correlation of an unresolved
 $5^{th}$ magnitude star with one pair of telescopes. For an actual 
measurement across one baseline, one must achieve a sensitivity corresponding 
to $ |\gamma_d|^2 \approx 0.5 $ and this time has to be at least quadrupled. 
(*) These original proposals for HESS and VERITAS are not yet being deployed,
 all others are in operation or under construction. 
(**) ``Next-Generation'' does not correspond to any specific project
or experiment but provides an indication of the major parameters of possible
future large arrays currently being studied.}
\label{chertel}
\end{table}
  
\begin{figure}[focplan]
\epsscale{.80}
\plotone{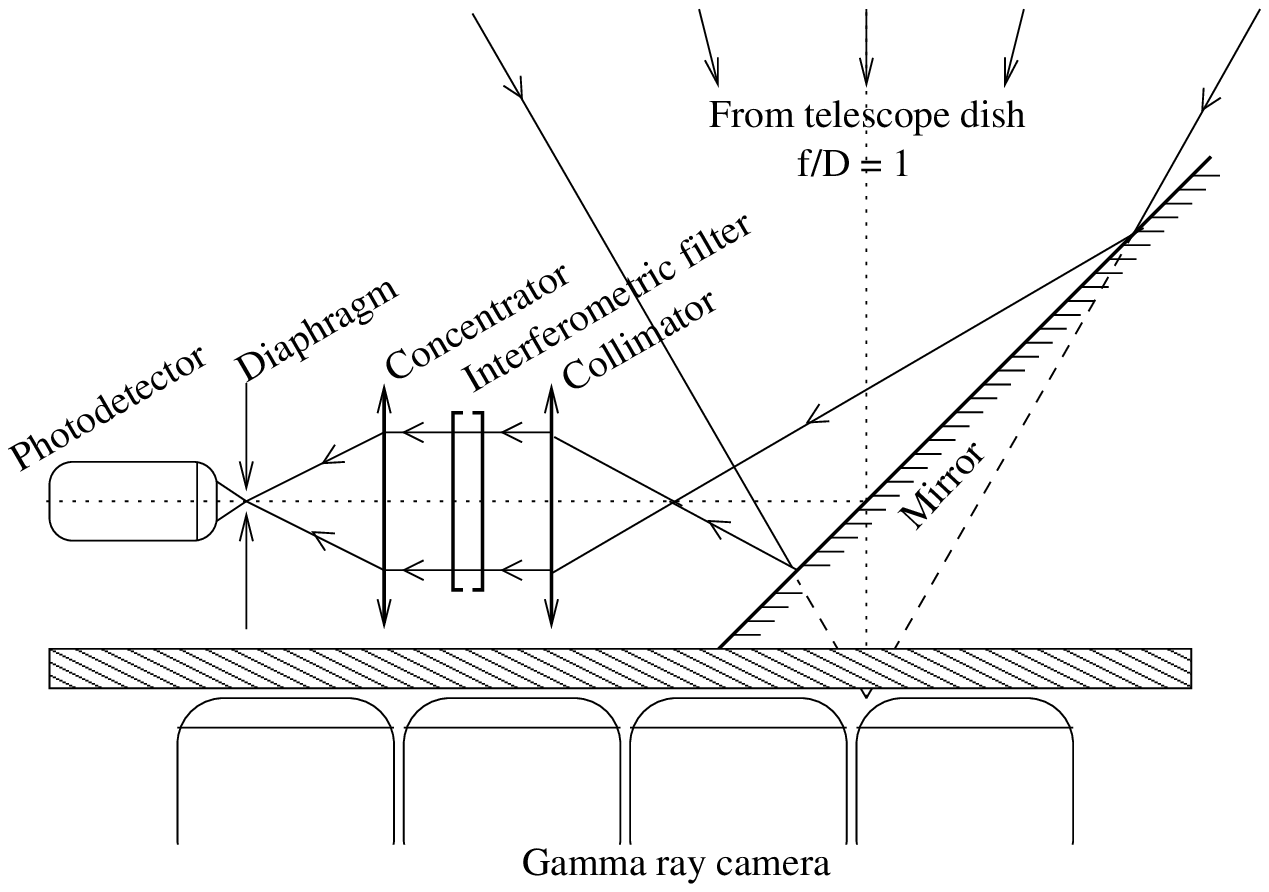}
\caption{The light from the telescope is redirected sideways by a mirror, it is 
then collimated, filtered and focused onto the photo-detector.}
\label{focplane}
\end{figure}

Converting an ACT telescope into an intensity interferometer element seems 
quite simple and inexpensive. A plate is mounted in front of the camera, 
protecting the photomultipliers used for $\gamma$-ray observations and serving 
as an optical bench for the interferometer elements. 
 A mirror at a $45^o$ angle from the telescope axis is used to redirect the 
light towards a lense used as a collimator. After passing through a narrow
bandwidth filter, the light can be concentrated on a photodetector as 
shown schematically in Figure \ref{focplane}.

At this point we have shown that the idea of using ACT arrays as intensity
interferometer seems promising and does not encounter any obvious practical
problems. ACT arrays have scientific programs which are incompatible with
interferometry observations. However, in practice, ACTs are only used for
$\gamma$-ray observation during Moonless nights. Stray light from the Moon 
should only slightly reduce the sensitivity of an intensity interferometer 
and telescope time allocation could be set according to the Moon visibility,
leaving almost half the night time available for interferometry
measurements. In the next section we discuss specifics and practical issues 
related to details of the implementation of an intensity interferometer using 
ACT arrays.
 
\section {Optics}

\subsection{Limitations due to the specific optical design of many ACTs}
Atmospheric Cherenkov telescopes are designed with an angular resolution  
optimized for 
Cherenkov images over the widest fields of view possible, with aperture ratios
close to unity. The solution adopted by HESS and VERITAS was to use the 
Davies-Cotton design \cite{Davies57}, which does not preserve isochrony. For 12\,m telescopes 
with aperture ratio close to one, this results in an effective 
bandwidth limitation close to 100\,MHz. With these telescopes, it is therefore 
not possible to improve sensitivity over the Narrabri intensity interferometer 
by utilizing the larger bandwidths nowadays available. The MAGIC and 
CANGAROO telescopes have parabolic mirrors which do not cause any time 
dispersion. For these telescopes, it would be possible to use much higher 
bandwidth, probably up to $\rm 1\,GHz$, which would result in a sensitivity 
improvement of 1.25 magnitudes which was not included in
table \ref{chertel}.   

\subsection{Consequences of the fixed positions of telescopes in ACT arrays}
At Narrabri, the two telescopes could be moved along tracks to keep the 
signals in time and maintain a fixed baseline as the star was tracked for 
long periods of time. In ACT arrays, telescopes are at fixed locations. This
has three consequences complicating the use of these telescopes in an
intensity interferometer.   

First, the signals from different telescopes have to be brought back in time
for the correlation to be measured. Programmable delays can be used to
accomplish this; an analog delay system with similar requirements has been
operated by the CELESTE experiment \cite{pare}. It should be noted that the
accuracy necessary for the timing corrections is fixed by the system
bandwidth. For the contribution from the highest frequency preserved by the
system to be degraded by less than 10\%, the timing must be controlled with an
accuracy better than 7\% of the corresponding period. For example, a $\rm \sim
100\,MHz$ system requires timing corrections to within $\rm \sim 0.7\,ns$.

Second, the length and orientation of the effective baseline between two 
telescopes depends on the position of the star in the sky. Stars with 
diameters several times larger than the $\rm 1.22 \lambda / d $ indicated in 
table \ref{chertel} could effectively be measured by making observations at 
lower elevation where baselines are effectively reduced. The baseline will 
also  change with time during the observation of any star. This can be taken 
into account at analysis time provided that intermediate results are stored 
regularly. This turns out to be useful as a continuous range of effective
baselines can be probed in one night, even with a single pair of
receivers. This is already used in radio and optical interferometry 
\cite{davis2005}. 

Third, in ACT arrays, there are no close pairs of telescopes with which we 
could obtain close to zero baseline measurements. The optimal distance between 
telescopes is set for $\gamma$-ray observations by the radial extension of the 
Cherenkov light pool produced by atmospheric high energy showers, which is in 
excess of $\rm 120\,m$. As a consequence, telescopes would not be placed at 
much less than $\rm 80\,m$ from each other. Zero baseline interferometer 
measurements are, however, desirable. 
It is  in fact possible to 
obtain a zero 
baseline measurement from single telescopes. This can be done by splitting 
the beam of collimated and filtered light. The correlation of the 
fluctuations in the two beams from one telescope then provides a zero baseline
measurement of the coherence. Each telescope could be used for zero baseline 
measurements and still be used in other baselines.

\subsection{Noise limitations associated with the field of view}
Cherenkov images of atmospheric showers of interest to very high energy
$\gamma$-ray astronomy are elongated with typical lengths of half a degree and
widths of a few tenths of a degree. The width of these images sets the ACT
pixel size to $0.1^\circ - 0.2^\circ$ and the optical system is typically
designed to produce a point spread function matching the pixel
size. Sensitivity estimates presented above were made assuming that the only
noise was due to the Poisson fluctuations in the light from the star being
measured. The finite size of the point-spread function makes it impossible to
avoid also having some amount of night sky background (NSB) light polluting
the star light. The NSB light collected by two telescopes does not produce any
correlation signal, but by increasing the shot noise it limits the sensitivity
that can be attained in a given observation time. If $\rm \sim 30\%$ of the
collected light is not coming from the star, the time necessary to attain a
given significance for the measurement of the correlation is increased by a
factor two. The continuum spectral density from a star of visual 
magnitude V is
approximately $5 \times \rm10^{-5} \,m^{-2}s^{-1}Hz^{-1} 2.5^{-V}$ while from
the NSB it is close to $5 \times \rm10^{-3} \,m^{-2} s^{-1} Hz^{-1} sr^{-1}$
\cite{ozlem}. From this, we see that with a  typical point spread function of
$0.05^\circ$ ($\rm \sim 6 \times 10^{-7} sr$ )  as in MAGIC, HESS or VERITAS, 
stars of magnitude 9.6 are the
dimmest that could practically be measured. If it were not for this
limitation, solar arrays used for atmospheric Cherekov $\gamma$-ray observation,
such as STACEE\cite{dhanna}, would be extremely attractive, providing several hundred
receivers each with a $\rm 37\,m^2$ light collecting area (this solution was
recently proposed by Ofir \& Ribak 2006b). However, the point spread function 
of these mirrors typically extends over close to half a degree, severely 
limiting the sensitivity that could be achieved if the array were used as an
interferometer.


\section {Electronics}

\subsection{The photo-detectors}
Since the seventies, developments in photo-detection have been dramatic. The two
most important aspects for us to consider are bandwidth and quantum
efficiency. Some modern photomultiplier tubes have bandwidths of more than 
1\,GHz. As already mentioned, in comparison to the 60\,MHz bandwidth used in
Narrabri, this corresponds to a sensitivity gain of more than one magnitude
provided the optical system has a compatible bandwidth. As for the quantum
efficiency, solid state photo-detectors seem attractive with quantum 
efficiencies of
more than 70\%. Unfortunately, for large enough photodiodes, the 
bandwidth is not as good as can be found in photomultipliers and, more 
importantly,  noise prevents them from having the
sensitivity required for stellar intensity interferometry. Photomultipliers
with GaASP photocathode, however, have quantum efficiency curves peaking
close to 50\%, which corresponds to a gain of more than half a magnitude if
compared to the 25\% quantum efficiency in the Narrabri interferometer.
Combining improvements in bandwidth and quantum efficiency since the 70s, the
observation time required to reach a given sensitivity can be halved. 

\subsection{Cable bandwidth and large baselines}
Analog signals could be brought to a central
location and then correlated using analog or digital techniques. In this
approach the bandwidth of the signal transmission
over several hundreds of meters is critical. Even broadband cable such as RG-6 gives more
than 4\,dB attenuation over 100\,m at 50\,MHz. Analog signal transmission bandwidth
can effectively limit the sensitivity of telescopes such as MAGIC or CANGAROO
which, in principle, have the capability of working with larger bandwidth. A
possibility for circumventing this problem is to use analog optical fiber
transmission. This is already being used in the MAGIC system for $\gamma$-ray
observation and the MAGIC collaboration has demonstrated the possibility of
transmitting analog signals with attenuations of $\rm 3\,dB/Km$ at $\rm 500\,MHz$ with a
dynamic range of $\rm 62\,dB $\cite{paneque2003}. Such a dynamic range is
sufficient to allow for the measurements over a range of more than 10
magnitudes without changing gains, and in practice does not constitute a
limitation.

Alternatively, it is possible to digitize the signals locally at each receiver
and have the digitized signals sent to a central station where they can be
combined to provide a measurement of their correlation. One difficulty of this
approach resides in synchronizing the two digitizers. The difficulty of
achieving synchronization of the digitizer depends on the digitization
rate. This approach presents the advantage of eliminating all risk of noise
contamination and signal attenuation in cables between the photodetectors and
the correlator.

\subsection{The correlator}
The correlator must integrate over time the product of the AC components of 
two signals. The DC components must also be measured in order to normalize the
correlation. Modern electronics make these tasks easier than it was in the 
60s. Again, two approaches are now possible: analog and digital. 

\begin{figure}[tabletop]
\epsscale{.80}
\plotone{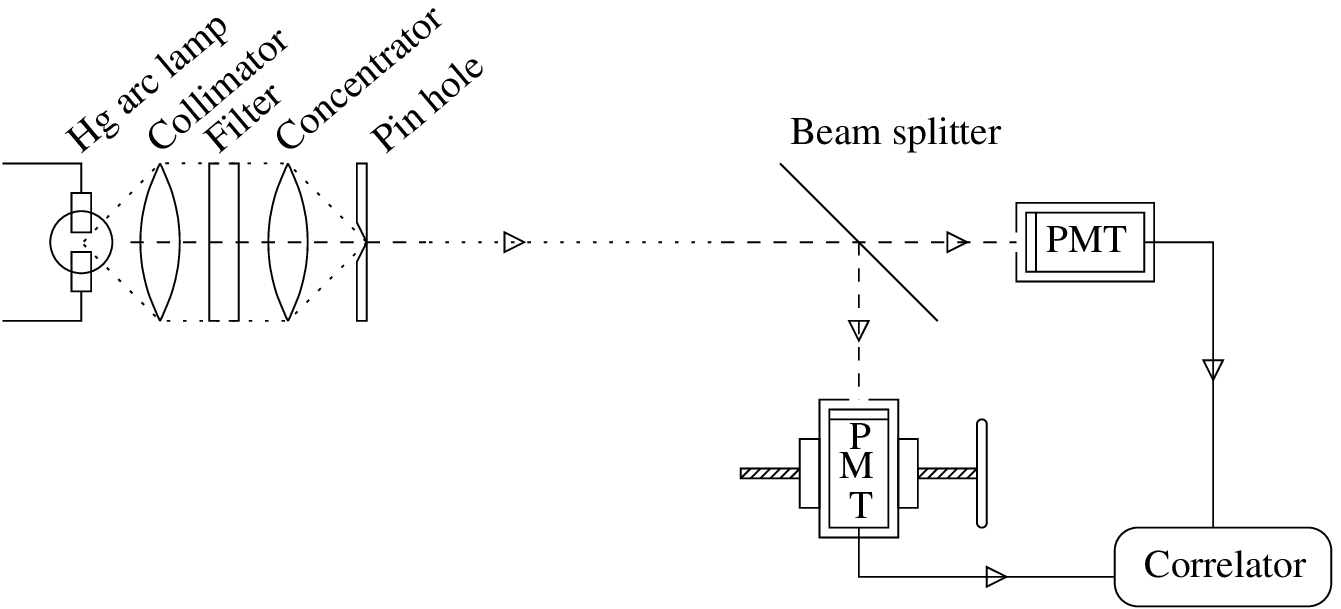}
\caption{The light from a mercury arc-lamp is filtered and focused on a 
pinhole to form an artificial star of known diameter. Two photo-detectors 
are virtually overlapped by means of a semi-reflective glass for the zero 
base-line measurement and the correlation is measured as a function of the 
lateral displacement of one with respect to the other.}
\label{tabletoptest}
\end{figure}

Signals could be sent through analog delays and, after proper amplification,
be collected by a double-balanced mixer, a component used in a variety of
radiofrequency applications which, in effect, forms the product of its two
analog inputs. The mixer output is received by a voltage integrator. This
method was successfully used in the intensity interferometer designed for
synchrotron X-ray beam diagnostics by Yang et al. (1994). The drift of signals
away from the zero point, which produces a false correlation, could be
removed using the same phase-switching technique originally used by
Hanbury Brown and Twiss.  Nowadays this could also be achieved by using a lock-in
amplifier as described in Tai et al. (2000). We are working on a table-top
experiment (Figure \ref{tabletoptest}) using this method.  

\begin{figure}[schem]
\epsscale{.50}
\plotone{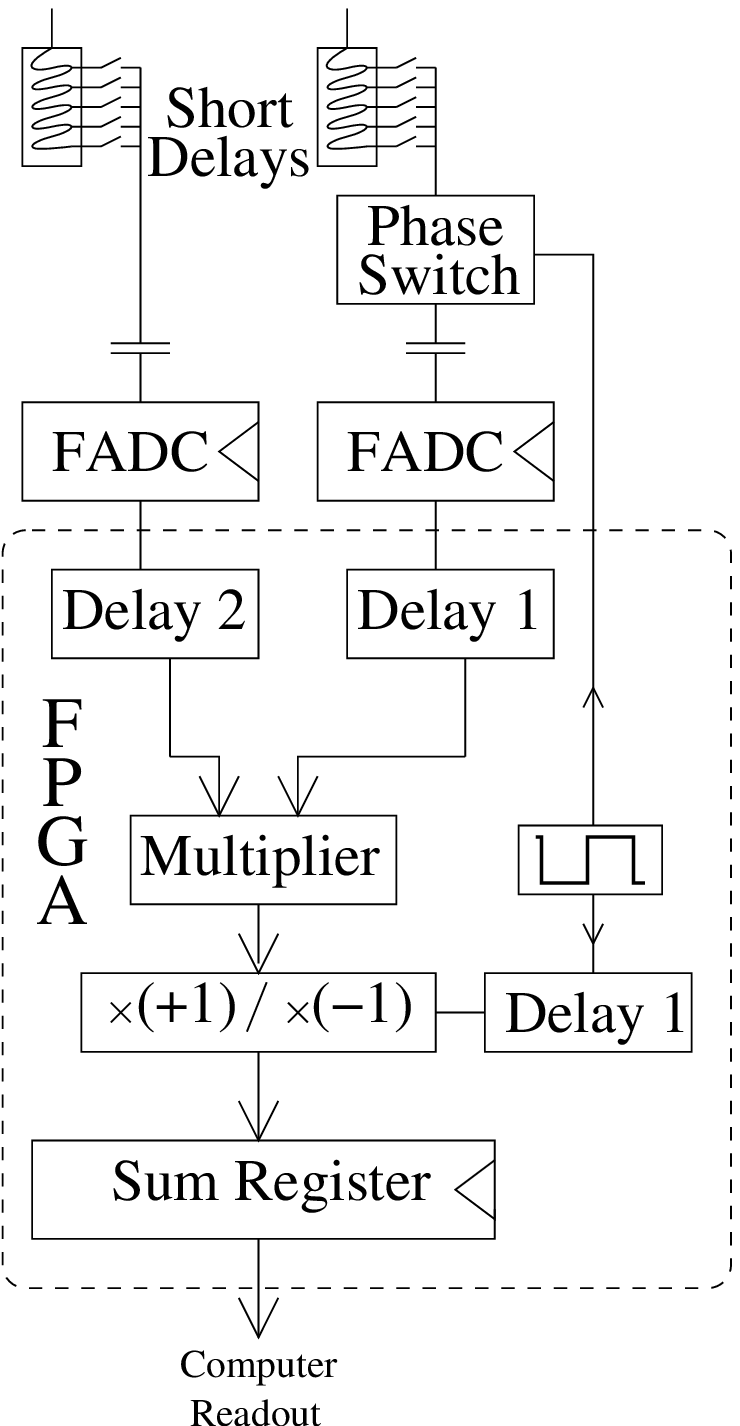}
\caption{Short programmable delay lines can be used to set the relative 
  timing of the two channels to a fraction of the digitization period. Large 
  time corrections can be done on the digitized samples. The phase of one of 
  the two signals is inverted with a 50\% duty cycle before digitization. 
  Digital samples are multiplied together. In order to demodulate the 
  correlation signal, the sign of the product is inverted during phases when 
  the phase of the analog signal is inverted. The result is accumulated in a 
  sum register.}
  \label{schem}
\end{figure}

The digital approach may follow two different paths. It is possible to, at 
least temporarily, store the digital data from each telescope and run an 
offline analysis to obtain the various correlations needed. The difficulty is the 
volume of data that has to be stored and manipulated (more than $\rm 200\,Mb$ 
per telescope and per second for a $\rm 100\,MHz$ bandwidth system). 
Alternatively, the digitized signals, once in a central location could be 
duplicated an indefinite number of times and sent to correlators implemented 
in Field Programmable Gate Arrays (FPGA). These could be programmed to bring the 
signals in time up to the digitization period before they are multiplied. 
This does however not eliminate 
the need for short analog delays. As we have seen, timing should be 
adjusted with an accuracy of a fraction of the period of the highest 
frequency which, in a digital system, could be the assimilated to the 
Niquist frequency.  The unavoidable 
offsets from the digitization would be responsible for drifting that would 
drown the correlation signal. This can be eliminated by phase switching. Figure \ref{schem} is a functional diagram 
presenting a possible implementation for a two channel system. 
The great advantage of a digital 
system implemented in an FPGA is that the algorithm can be modified at will. 
For example it could be adjusted to filter-out noise and reduce
spurious correlations. We are starting to develop such a digital correlator 
which will be tested on the table-top experiment described in Figure 
\ref{tabletoptest}. 

\section{Conclusions}

We have previously presented the possibility of implementing an intensity
interferometer using atmospheric Cherenkov telescope arrays
\cite{lebohec05a,lebohec05b}. Here, we have shown that existing atmospheric
Cherenkov telescopes could be used as the receivers of intensity
interferometers to study stars of magnitude as large as 5 without taking
advantage of the technological developments since the time of the Narrabri
interferometer. Higher sensitivities are achievable with the faster and more
stable electronics available today. These improvements, however, will not
affect the sensitivity limit arising from the quality of the receiver optics.
We have seen that existing ACTs, with a point spread function of $\sim0.05^o$
are restricted to objects of magnitudes smaller than $\rm \sim 9.6$ with a
$\rm \sim 5$ hour integration. The optical requirements for interferometric
studies could be taken into account in the design of future arrays of larger
Cherenkov telescopes to make them more effective.

We have considered the constraints and technical difficulties associated with
such a project and proposed some possible technical solutions without finding
major difficulties. We conclude that available telescope arrays can, in
principle, be used for the interferometric study of stars at wavelengths
shorter than $\rm 400\,nm$ with baselines close to $\rm 200\,m$. This would
give access to angles as small as $\rm
\sim0.1\,mas$. Figure~\ref{ang_diam_veritas} shows that, for example, the HESS
or VERITAS baselines allow for measurements of the correlation curve for stars
with angular sizes of $\rm \sim0.5\,mas$, corresponding to a star the same
radius as the sun at a distance of $\rm \sim20\,pc$. Future Cherenkov
telescope arrays being discussed extend over $\rm \sim 1\,km$ and could be
used to measure angles ten times smaller. One should then question the
importance and role intensity interferometry could play in the coming years.

We note that the measurements provided by the Narrabri intensity 
interferometer of 32 stars, in the spectral range O5f to F8 \cite{brown1974}, 
are still relevant and competitive today \cite{kervalla} although the 
instrument ceased taking measurements as long ago as 1972. 
The most useful observations which could be made with a modern-day intensity 
interferometer are guided mainly by the unique strengths of the 
technique.  The relative ease of long-baseline measurements allows for very 
high angular resolution.  A further consideration is the ability to make 
measurements at shorter wavelengths than those usually probed by Michelson 
devices. Finally, long exposures using ground-based $\gamma$-ray telescopes 
under moonlight are effectively free of competition for observing time, 
allowing the possibility of very long-term monitoring of variable sources.

\begin{figure}[stars]
\plotone{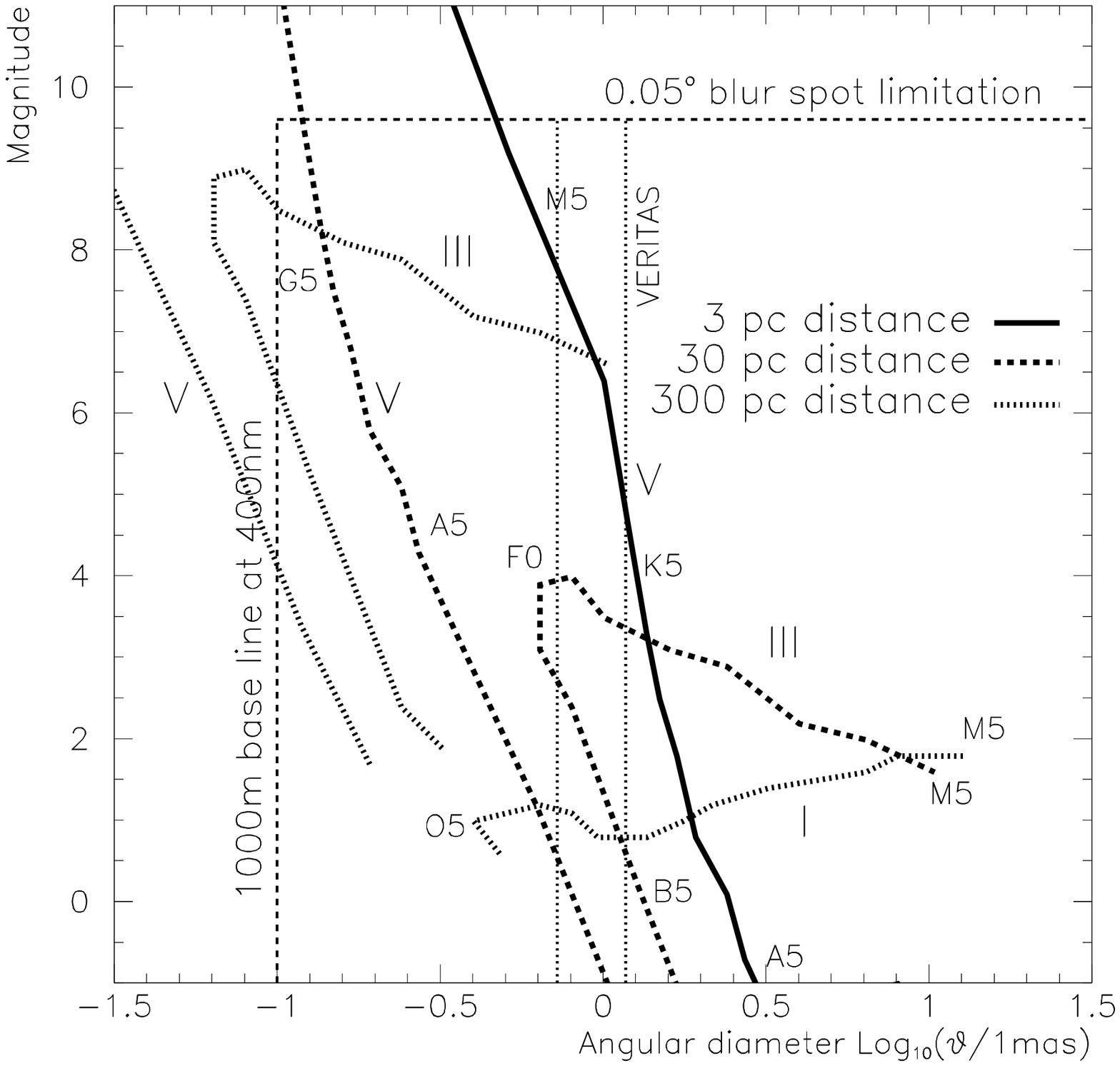}
\caption{Visual magnitude - angular diameter relationship for 
the main sequence, the giant and the super-giant branches for distances of 
3\,pc, 30\,pc and 300\,pc. The magnitude limitation resulting from a $0.05^o$ 
optical point-spread function is indicated, as well as the VERITAS baselines 
as an example. Future atmospheric Cherenkov telescope arrays might provide 
baselines of 1\,km.}
\label{stars}
\end{figure}

The scientific potential of optical interferometry in general is outlined in
e.g. \cite{monnier}. Figure \ref{stars} illustrates the capabilities of a
ground-based $\gamma$-ray telescope array at measuring stellar angular
diameters. Knowledge of the angular diameter of stars is important for the
understanding of their fundamental properties. Angular measurements are
affected by the limb darkening in a way that depends on the wavelength
\cite{mozurkewich}. Intensity interferometers, with their capability of
working over a range of short visible wavelengths, could contribute to the 
measurements of the stellar diameter wavelength dependence. The long 
base-lines offer the
possibility of measuring main sequence late-type stars for which data is still
missing\cite{kervalla}. Measurements of Cepheid stars can be used to obtain
improved calibration of the period-luminosity relation \cite{kervalla2006}.
Arrays of more than two telescopes could also be used to measure the
oblateness of rapid rotators, providing further parameter constraints on
stellar evolution models\cite{vanbelle2006}. Future arrays with many base
lines simultaneously available could even measure finer details on stellar
surfaces as in \cite{ young2000}. The observation of spectroscopic binaries is
also of interest; resolving the components of these systems would determine
all of the system parameters, including the respective members' masses
\cite{davis2005}. It is also simple, with an intensity interferometer, to
select emission lines so as to study the circumstellar environment. The NPOI
was used recently to study $\rm H-\alpha$ emitting regions in Be stars
$\gamma$ Cassiopeiae and $\phi$ Persei with base lines of $\rm 40\,m$ and
less. Using an ACT intensity interferometer with an $H-\beta$ emission line
filter could possibly reveal details ten times
smaller\cite{tycner2006}. Objects such as $\zeta$-Tauri stars
\cite{tycner2004}, Wolf-Rayet stars, Luminous-Blue-Variables
\cite{chesneau2003} or even novae \cite{quirrenbach} could also be monitored
at short wavelength with unprecedented interferometric baselines.

In summary, atmospheric Cherenkov telescope arrays provide an inexpensive and
practical way by which to implement a modern intensity interferometer. The
technical difficulties are few, and the successful operation of an intensity
interferometer system on currently operating ACT arrays would significantly
increase the scientific output of these instruments. The experience gained
from operating such a system would also allow for direct input, relevant to
interferometry measurements, into the design of next generation systems.

\acknowledgments{ 
SLB acknowledges support from NSF grant \#58500944 and University Of 
Utah Research Foundation. JH acknowledges support from PPARC. We are grateful 
to David Kieda, Jeremy Smith , Jeter Hall and the members of the VERITAS 
collaboration for many useful discussions. We would also like to thank Micah 
Kohutek for setting up the table-top test bench and Emmet Valfredini for his
technical assistance.}

\end{document}